\documentclass[useAMS,usenatbib,a4paper,referee]{mn2e}
\usepackage{graphicx}
\usepackage{amsmath}
\usepackage{txfonts}
\usepackage{mathrsfs}

\def\aap{A\&A}
\def\apjl{ApJ}

\def\apjs{ApJS}
\def\apj{ApJ}

\newcommand{\be}{\begin{equation}}
\newcommand{\ee}{\end{equation}}
\newcommand{\bea}{\begin{eqnarray}}
\newcommand{\eea}{\end{eqnarray}}

\title[Epoch of Reionization]{The Epoch of Reionization in the $R_h=ct$ Universe}
\author[Fulvio Melia and Marco Fatuzzo]{Fulvio Melia$^{1}$\thanks{John Woodruff Simpson 
Fellow. E-mail: fmelia@email.arizona.edu} and Marco Fatuzzo$^{2}$\thanks{E-mail: fatuzzo@xavier.edu}\\
\null$^{1}$Department of Physics, The Applied Math Program, and Department of Astronomy, 
The University of Arizona, AZ 85721, USA\\
\null$^{2}$Physics Department, Xaiver University, Cincinnati, OH 45207}

\begin{document}

\date{}

\pagerange{\pageref{firstpage}--\pageref{lastpage}} \pubyear{2013}

\maketitle

\label{firstpage}

\begin{abstract}
The measured properties of the epoch of reionization 
(EoR) show that reionization probably began around 
$z\sim 12-15$ and ended by $z=6$. In addition, a 
careful analysis of the fluctuations in the cosmic 
microwave background indicate a scattering optical 
depth $\tau\sim 0.066\pm0.012$ through the EoR. In 
the context of $\Lambda$CDM, galaxies at intermediate 
redshifts and dwarf galaxies at higher redshifts now 
appear to be the principal sources of UV ionizing
radiation, but only for an inferred (ionizing) escape 
fraction $f_{ion}\sim 0.2$, which is in tension with 
other observations that suggest a value as small as 
$\sim 0.05$. In this paper, we examine how reionization 
might have progressed in the alternative 
Friedmann-Robertson Walker cosmology known as the 
$R_{\rm h}=ct$ Universe, and determine the value of 
$f_{ion}$ required with this different rate of 
expansion. We find that $R_{\rm h}=ct$ accounts 
quite well for the currently known properties of 
the EoR, as long as its fractional baryon density 
falls within the reasonable range $0.026\lesssim 
\Omega_b\lesssim 0.037$. This model can also fit 
the EoR data with $f_{ion}\sim 0.05$, but only if 
the Lyman continuum photon production is highly 
efficient and $\Omega_b \sim 0.037$. These results 
are still preliminary, however, given their 
reliance on a particular form of the star-formation 
rate density, which is still uncertain at very high 
redshifts. It will also be helpful to reconsider 
the EoR in $R_{\rm h}=ct$ when complete structure
formation models become available.
\end{abstract}


\section{Introduction}
In the standard model of cosmology, the Universe entered the so-called
``dark ages" soon after recombination, at cosmic time $t\sim 380,000$ years, 
initiating a period that ended only when stars and galaxies began forming
some 400 Myr later (Barkana \& Loeb 2001; Bromm \& Larson 2004). It is thought that 
during the ensuing $\sim 500$ Myrs, Lyman continuum radiation from early galaxies 
and emerging active galactic nuclei (AGNs) reionized the expanding gas, producing 
a fully ionized intergalactic medium (IGM) by redshift $z\sim 6$. This termination 
point is well established observationally, e.g., through the Gunn-Peterson absorption 
measured in high-redshift quasars, whose spectra reveal that hydrogen was highly 
ionized by $t\sim 1$ Gyr (e.g., Songaila 2004; Fan et al. 2006). And while the 
precise time (or redshift) at which the Epoch of Reionization (EoR) began is not 
as well established, indications from, e.g., the cosmic microwave background (CMB) 
polarization data, are that it  probably began no later than $z\sim 10-15$ 
(Jarosik et al. 2011; Hinshaw et al. 2013).

But though it is generally understood that the IGM was ionized by the integrated
UV field from AGNs and star-forming galaxies (Miralda-Escude \& Ostriker 1990; 
Haardt \& Madau 1996), the relative contributions from them, or even which 
dominated the UV emission first or last, are issues that have not yet been fully
resolved. Recent work constraining HI reionization by high-redshift AGNs, based on 
observed limits to the unresolved X-ray background (Haardt \& Salvaterra 2015), 
suggests that to avoid over-producing the X-ray signal measured at $z=0$, such 
quasars could not have been responsible for more than $\sim 13\%$ of the HII 
filling factor by $z\sim 6$. This conclusion comes with an important caveat,
however, in that other AGNs may have been present, but were heavily obscured 
and therefore too faint to be seen in X-rays. Absent such a population, 
the observational evidence for the dominant source of ionizing radiation at 
$z\gtrsim 6$ is beginning to favor a combination of bright galaxies at 
intermediate redshifts (e.g., Madau et al. 1999; Gnedin 2000; Wyithe \& 
Loeb 2003; Meiksin 2005; Trac \& Cen 2007; Faucher-Gigu\`ere et al. 2008; 
Gilmore et al. 2009; Vanzella et al. 2010) and dwarf galaxies ($M\lesssim 
10^9\;M_\odot$) filling the distribution out to $z\gtrsim 10-12$ (e.g., 
Robertson \& Ellis 2012; Robertson et al. 2015).

The central question then becomes whether or not the UV radiation leaking 
out of these galaxies is sufficient to complete the reionization process
by $z\sim 6$. Direct measures of the ionizing Lyman continuum flux from
galaxies at $z\gtrsim 5$ are not feasible due to the saturated hydrogen
absorption by the IGM. At lower redshifts ($z\sim 3-4$), estimates are possible,
but the inferred values seemingly depend on specific assumptions and mode of
analysis. Nestor et al. (2011) and Mostardi et al. (2013) have concluded
that the escape fraction $f_{ion}$ may be as high as $\sim 10-15\%$, though Vanzella
et al. (2012) have questioned these numbers on the basis of significant
contamination by foreground, low-redshift interlopers. An alternative
approach, using spectroscopic and very deep broadband and narrowband
imaging, suggests that $f_{ion}$ may be as small as $\sim 5\%$ (Vanzella et al.
2010; Boutsia et al. 2011). The escape fraction from local galaxies
may be even smaller than this, perhaps on the order of $\sim 1\%$.

However, these limits don't necessarily apply to the dwarf galaxies,
which may have larger escape fractions (e.g., Fontanot et al. 2014). The
dominant contributors to the cosmic reonization may therefore be these
fainter galaxies extending out to $z\sim 10-12$ (Ferrara \& Loeb 2013;
Wise et al. 2014; Yue et al. 2014). The most recent work on this possibility 
(Robertson et al. 2015), based on the actual measured star-formation rate (SFR)
history (Madau \& Dickinson 2014), has concluded that an escape fraction
$f_{ion}\sim 20\%$ is required in order to match the observed onset and
duration of the EoR. But is such a large fraction realistic?  Future 
work with lensing galaxy clusters to measure the Lyman continuum flux 
released into the IGM by gravitationally lensed, intrinsically faint 
galaxies may soon provide a better answer (see, e.g., Vanzella et al. 2012, 
Ishigaki et al. 2014).

All these uncertainties leave open the possibility that as the accuracy
of the measurements improves, particularly with regard to $f_{ion}$, it
may be difficult within the context of $\Lambda$CDM to reconcile the 
observed properties of the EoR with the known sources of UV ionizing
radiation. Perhaps the problem is not so much the lack of adequate
sources but, rather, the amount of time available within the interval
$6\lesssim z \lesssim 15$ for the reionization to have been completed. In
other words, if this tension persists, it may be an indicator that
the redshift-age relationship predicted by the standard model
is not consistent with the properties of the EoR. 

A precedent for such a proposal has already been set by the apparent
early emergence of  supermassive black holes at $z\gtrsim 6$ (Melia 2013a) 
and $\sim 10^9\;M_\odot$ galaxies at $z\sim 10-12$ (Melia 2014a), the very 
objects now thought to be responsible for the reionization. In the
concordance model, the Universe was simply not old enough by $z\sim 6$
and $z\sim 12$, respectively, for such objects to have formed. One of the
principal goals of this paper is therefore to examine how the requirements
on $f_{ion}$ might change when the onset and duration of the EoR are
matched to the predictions of an alternative FRW cosmology known as the 
$R_{\rm h}=ct$ Universe (Melia 2007, 2013b, 2015b; Melia \& Shevchuk 2012).

In recent years, we have carried out many comparative tests between 
$R_{\rm h}=ct$ and $\Lambda$CDM, showing that the data tend to favor the 
former with a likelihood $\sim 90\%$ versus $\sim 10\%$, according to the 
Akaike (AIC) and Bayesian (BIC) Information Criteria (see, e.g., Wei et al.
2013; Melia \& Maier 2013; Melia 2014b, 2015a; Wei et al. 2014a, 2014b; Wei et al. 
2015a, 2015b; Melia et al. 2015). Quite significantly in the 
context of this paper, the $R_{\rm h}=ct$ cosmology completely mitigates 
the tension created by the otherwise early appearance of high-redshift quasars 
and dwarf galaxies, because in this cosmology the EoR started at $\sim 800$ 
Myr ($z\sim 15$) and ended at $t\sim 1.89$ Gyr (i.e., $z\sim 6$), providing 
just the right amount of time for these structures to have grown according to
standard astrophysical principles as we know them (Melia 2013a, 2014).

In this paper, we will take as our starting point the most recent constraints
established for the sources of UV ionizing radiation, the measured star-formation
rate as a function of redshift, and current limits on the optical depth through
the IGM, and compare in detail the various contributions to the ionized filling 
factor in the $R_{\rm h}=ct$ and $\Lambda$CDM cosmologies. In our analysis, 
we include both models because significant progress has already been achieved 
in tracking the EoR in $\Lambda$CDM, so it should be easier to understand 
the differences between the two expansion scenarios. In the relevant redshift 
range $6\lesssim z\lesssim 15$, these models differ not only in their predicted 
age-redshift relationship, but also in their comoving volumes and corresponding 
densities. So to fully appreciate the different outcomes, particularly with 
regard to the required value of $f_{ion}$, we will consider each effect 
separately, and then track the overall ionized fraction as a function of 
redshift. We begin with an accounting of these model differences
in \S~2, and provide an overview of reionization in \S~3.  We solve
the governing equations and apply observational constraints to limit 
our paramter space in \S~4 and \S~5, and provide a discussion of our
results in \S~6.  A summary of our conclusions is presented in \S~7.

\section{Principal Differences between $\Lambda$CDM and $R_{\rm h}=ct$}
A proper analysis of the history of reionization in the IGM requires
knowledge of both the ionization rate and recombination time as functions 
of $t$. The former is primarily dependent on the star-forming galaxy density, 
while the latter depends on the physical conditions in the IGM. Both of
these quantities are cosmology dependent, so we begin by reviewing 
the relevant differences between these two models. Throughout this work, we 
adopt the most recent {\it Planck} (Ade et~al.\ 2014) parameters for 
$\Lambda$CDM: $H_0=67.74$ km s$^{-1}$ Mpc$^{-1}$, $\Omega_m = 0.309$, 
$\Omega_b h^2= 0.02230$, $Y_p = 0.2453$, and $w_{de} = -1$, where $Y_p$ is
the Helium fraction by mass and $w_{de}$ represents the dark-energy
equation-of-state. The ratios $\Omega_m\equiv \rho_m(t_0)/\rho_c$ 
and $\Omega_b\equiv \rho_b(t_0)/\rho_c$ are defined in terms of today's
(luminous plus dark) matter and baryon densities, respectively, and 
the critical density
\be
\rho_c \equiv {3 H_0^2 \over 8 \pi G} = 1.8785 \times 10^{-29}\,h^2\;{\rm g}\,{\rm cm}^{-3}\,.
\ee
The CMB fluctuations have not yet been fully analyzed in the context of 
$R_{\rm h}=ct$ (but see Melia 2014b, 2015), so the parameters in this 
model have not yet been optimized in this fashion. To keep the comparison as 
simple as possible, however, we will assume the same values for $H_0$ 
and $Y_p$, since these have been established observationally using
several means. We will discuss how the other parameters differ below. 

\begin{figure}
\center{\includegraphics[scale=0.6,angle=0]{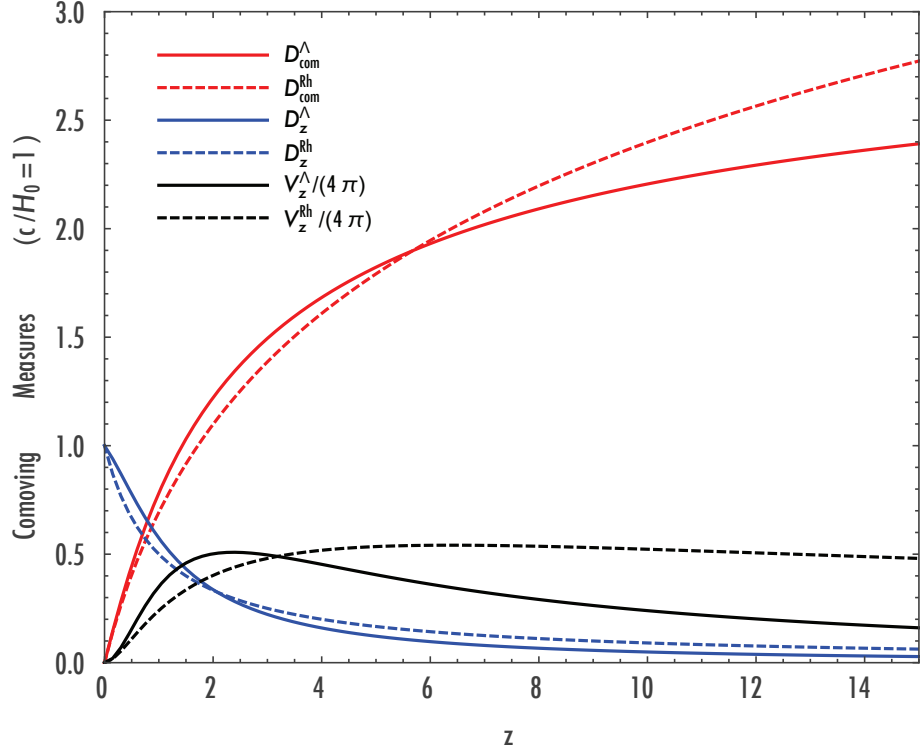}
\caption{A comparison of various distance measures as functions of $z$ in
$\Lambda$CDM and the $R_{\rm h}=ct$ Universe.}}
\end{figure}

\subsection{Comoving Distance and Volume}
The comoving distances in $\Lambda$CDM and $R_{\rm h}=ct$ are given, respectively, 
by the expressions
\be
D_{\rm com}^{\Lambda} = {c\over H_0} \int_0^z {du\over\sqrt{\Omega_m (1+u)^3 + 
\Omega_r (1+u)^4 + \Omega_\Lambda (1+u)^{3+3w_{de}}}}\;,
\ee
and
\be
D_{\rm com}^{R_{\rm h}} = {c\over H_0} \ln (1+z)\;.
\ee
From these, it is straightforward to calculate $D_z\equiv d D_{\rm com}/dz$
and the comoving differential volume $V_z\equiv d V_{\rm com} /dz= 4 \pi 
D_{\rm com}^2 \,d D_{\rm com} / dz$, all of which are shown as functions of
redshift in figure~1.

\subsection{The Age-Redshift Relationship}
The age of the Universe at redshift $z$ in $\Lambda$CDM is 
\be
t^{\Lambda}(z) = {1\over H_0} \int_z^\infty {du\over\sqrt{\Omega_m (1+u)^5 
+ \Omega_r (1+u)^6 + \Omega_\Lambda (1+u)^{5+3w_{de}}}}\;.
\ee
The corresponding expression in $R_{\rm h}=ct$ is 
\be
t^{R_{\rm h}}(z) = {1\over H(z)}\;,
\ee
where $H(z)=H_0(1+z)$. The quantities $t^{\Lambda}$ 
and $t^{R_{\rm h}}$ are plotted in figure~2. In addition, we show in figure~3 
the ratios $R_t(z)\equiv t^{R_{\rm h}}/t^{\Lambda}$ and $R_V(z)\equiv V_z^{R_{\rm h}}
/V_z^{\Lambda}$ as functions of $z$. These are among the most important
influences affecting the ionization rate and recombination time in these two cosmologies,
which we will describe in greater detail shortly. 
\begin{figure}
\center{\includegraphics[scale=0.6,angle=0]{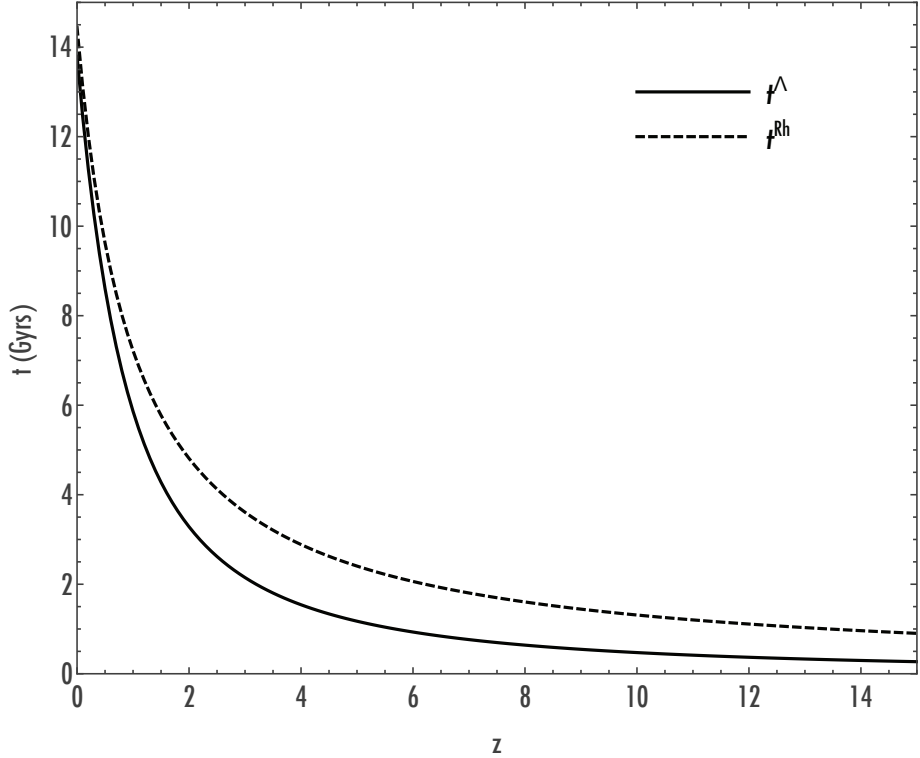}
\caption{The age-redshift relationship for $\Lambda$CDM and the $R_{\rm h}=ct$ Universe. }}
\end{figure}

\begin{figure}
\center{\includegraphics[scale=0.6,angle=0]{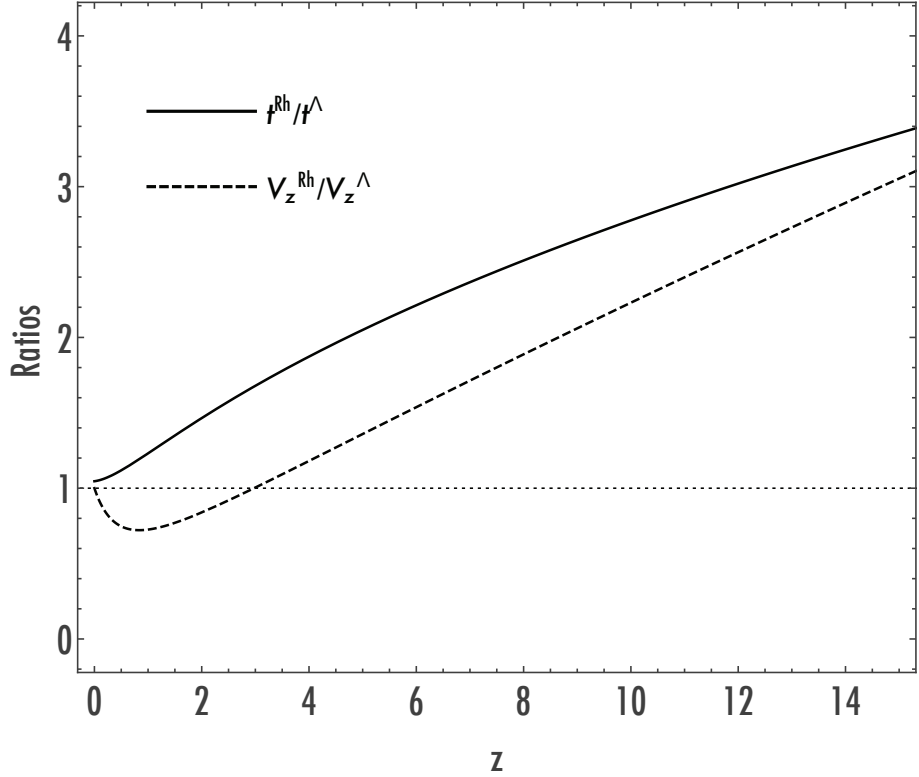}
\caption{Ratio of differential volumes $R_V$ and ratio of ages $R_t$
as functions of $z$ in $\Lambda$CDM and $R_{\rm h}=ct$.}} 
\end{figure}

\subsection{Baryon density}
Certainly up to redshift $z\sim 15$, it is safe to assume in $\Lambda$CDM that matter
evolves independently of the other components in the cosmic fluid, specifically
radiation and dark energy, though some questions have been raised concerning the 
possible instability, or creation, of dark matter. One typically takes the simplest
approach in this regard, to ignore such effects, and assume that both $\rho_m$ and
$\rho_b$ scale inversely with proper volume in this model, meaning that
\begin{eqnarray}
\rho_m^\Lambda&=&\Omega_m\rho_c(1+z)^3\;,\nonumber \\
\rho_b^\Lambda&=&\Omega_b\rho_c(1+z)^3\;.
\end{eqnarray}
The corresponding comoving densities $\bar{\rho}_m^\Lambda\equiv\Omega_m\rho_c$ and 
$\bar{\rho}_b^\Lambda=\Omega_b\rho_c$ are constant in $z$. The comoving Hydrogen number 
density is therefore given by the expression
\be
\bar{n}_H^{\Lambda} = {(1-Y_p) \Omega_b \rho_c \over m_H} = 
1.89 \times 10^{-7}\,{\rm cm}^{-3}\,.
\ee
The comoving number density of electrons $\bar n_e^{\Lambda} = f_e Q  \bar{n}_H^{\Lambda}$  
depends on the ionization state of the medium, where $Q$ is the ionization fraction of Hydrogen, 
and $f_e$ is a correction factor that accounts for the ionization of Helium.  For simplicity, we assume that 
Helium is either singly ionized or doubly ionized, so that
\be
f_e = 1+\xi {Y_p\over 4(1-Y_p)}\,,
\ee
where we take $\xi = 1$ for $z > 4$ and $\xi = 2$ for $z \le 4$ (Kuhlen \& Faucher-Gingu\'ere 2012).

The situation is a little different in the $R_h = ct$ Universe. The overall
dynamics in this cosmology has been tested with a wide range of observations,
from cosmic chronometers (Melia \& Maier 2013) and Type Ia SNe (Wei et al. 2015a)
in the local Universe, to young quasars (Melia 2013a) and the CMB (Melia 2014b, 2015a)
at high redshifts, but the detailed behavior of individual components in the cosmic 
fluid is only now beginning to be studied, using the kind of approach discussed
in this paper. The reason for this dichotomy is that, unlike $\Lambda$CDM in which
the expansion dynamics can be surmised only from the properties of the individual
constituents, the principal constraint in $R_{\rm h}=ct$ is the zero active mass 
condition, $\rho+3p=0$, where $\rho$ and $p$ are, respectively, the total energy
density and pressure in the cosmic fluid (Melia 2007, 2015b; Melia \& Shevchuk 2012).
For many applications, particularly when it comes to calculating 
the expansion rate and other observable quantities, such as the luminosity distance
and the redshift dependent Hubble constant $H(z)$, one does not need to know the 
detailed makeup of this fluid, since all of its components must together always 
produce a total equation-of-state $p=-\rho/3$, for which 
\begin{equation}
\rho^{R_{\rm h}}=\rho_c(1+z)^2\;.
\end{equation}

When the evolution of individual components is needed (as is the case here),
several conservation laws and reasonable assumptions delimit their behavior. 
At least in the local Universe and within the EoR, 
it is reasonable to assume (1) that the radiation evolves independently of 
matter and dark energy; (2) that (baryonic plus dark) matter exerts an insignificant 
pressure compared to radiation and dark energy; and (3) that the equation-of-state
parameter $w_{\rm de}\equiv p_{\rm de}/\rho_{\rm de}$ for dark energy is
constant. (This is not a requirement, but appears to be the simplest assumption
one can make.) 

For the baryon number, however, the situation is less clear.
Certainly, baryon number appears to be conserved in most interactions of the
standard model, but there are important exceptions, such as the chiral anomaly
(e.g., White 2004). Examples of this include sphaleron solutions to the electroweak 
field equations (e.g., Arnold \& McLerran 1987), involved in processes that violate 
baryon (and lepton) number conservation. But these are thought to be rare in the local 
Universe; they might have been much more common in the more extreme physical 
conditions prevalent in the early Universe, where sphalerons would have converted 
baryons to antileptons and antibaryons to leptons. Baryon number conservation
might also have been violated in grand unified theories, which could lead, e.g., 
to proton decay. 

Insofar as the EoR is concerned, we will assume that this period in the evolution
of the Universe was sufficiently far removed from the conditions in which baryon
number conservation would have been violated, so that the baryon density remained 
constant in the comoving frame. (This also assumes that baryons have no additional
interactions during the EoR with, e.g., the dark-energy field, when the standard
model of particle physics is extended. Otherwise, these calculations will almost
certainly have to be redone.) Therefore, the expression for $\bar{n}_H^{R_{\rm h}=ct}$
is identical to that for $\bar{n}_H^\Lambda$ in Equation~(7) (though the fitted values
for, e.g., $\Omega_b$ could be different). For the other quantities, we work with the 
following simultaneous equations describing the pertinent physics at redshifts $z\lesssim 15$:
\begin{equation}
\rho_{\rm de}+\rho_{\rm b}+\rho_{\rm d}+\rho_{\rm r}=\rho_{\rm c}(1+z)^2\;,
\end{equation}
\begin{equation}
w_{\rm de}\rho_{\rm de}+{1\over 3}\rho_{\rm r}=-{1\over 3}\rho_{\rm c}(1+z)^2\;,
\end{equation}
\begin{equation}
\rho_{\rm b}=\Omega_{\rm b}\rho_{\rm c}(1+z)^3\;,
\end{equation}
and
\begin{equation}
\rho_{\rm r}=\Omega_{\rm r}\rho_{\rm c}(1+z)^4\;.
\end{equation}
In these expressions, $\rho_{\rm d}$ is the energy density of dark matter,
defined by the equation
\begin{equation}
\rho_{\rm m}=\rho_{\rm b}+\rho_{\rm d}\,,
\end{equation}
and $\rho_{de}$ and $\rho_r$ are, respectively, the dark energy and radiation energy densities,
which are scaled analogously to $\Omega_m$ and $\Omega_b$ to produce the quantities $\Omega_{de}$ 
and $\Omega_r$ appearing below. These equations are easily solved to produce the evolution in 
$\rho_{\rm de}$ and $\rho_{\rm d}$ with redshift, complementing Equations~(12) and (13) for 
the other densities:
\begin{equation}
\rho_{\rm de}\approx -{1\over 3w_{\rm de}}\rho_{\rm c}(1+z)^2\left[1+\Omega_{\rm r}
(1+z)^2\right]\;,
\end{equation}
and
\begin{equation}
\rho_{\rm d}\approx \rho_{\rm c}(1+z)^2\left[2-\Omega_{\rm b}(1+z)-
{3w_{\rm de}-1\over 3w_{\rm de}}-{3w_{\rm de}-1\over 3w_{\rm de}}
\Omega_{\rm r}(1+z)^2\right]\;.
\end{equation}

Today's CMB temperature ($T_0\approx 2.72$ K) translates into a normalized 
radiation energy density $\Omega_{\rm r}\approx 5\times 10^{-5}$. Therefore, 
$w_{\rm de}$ must be $\sim -1/2$ in order to 
produce a partitioning of the constituents in line with what we see in the 
local Universe. With this value, 
\begin{equation}
\Omega_{\rm de}\approx -{1\over 3w_{\rm de}}\approx {2\over 3}\;,
\end{equation}
while
\begin{equation}
\Omega_{\rm m}\approx {1+3w_{\rm de}\over 3w_{\rm de}}\approx {1\over 3}
\end{equation}
where, of course, $\Omega_{\rm m}=\Omega_{\rm b}+\Omega_{\rm d}$.
Therefore, according to Equation~(16), $\rho_{\rm d}(z)\rightarrow 
0$ at $(1+z)\approx 15.6$. At this redshift, which we will call $z_*$, 
a dark-energy equation of state parameter $w_{\rm de}=-1/2$ would yield 
$\rho_{\rm de}\sim 0.68\,\rho(z_*)$, $\rho_{\rm m}= \rho_{\rm b}\sim 0.31\,
\rho(z_*)$, and $\rho_{\rm r}\sim 0.01\,\rho(z_*)$. The overlap of $z_*$ 
with the redshift at which the EoR is thought to have started may simply
be coincidental; it's not at all obvious why these two should be linked. 
On the other hand, it might be interesting to speculate on possible
physical reasons (beyond the standard model) for such a correlation,
though this kind of probe lies beyond the scope of the present paper. 

The data we are considering in this paper do not tell us much about
what is happening beyond $z_*$, so long as the medium at these high
redshifts is neutral (up to recombination). However, we point 
out for future reference
that, in $R_{\rm h}=ct$, a redshift $(1+z_*)=15.6$ corresponds to a 
cosmic age $t_*\equiv 1/H_0(1+z_*)\approx 950$ Myr, assuming a Hubble
constant $H_0=67.74$ km s$^{-1}$ Mpc$^{-1}$. Several of the above 
expressions and assumptions may not be valid at times earlier than
this. We do know that baryon number cannot be conserved during this
epoch, because otherwise $\rho_{\rm b}/\rho\rightarrow 1$ well short
of the big bang (from Equations~9 and 12). It is likely that at 
these early times the Universe may have been dominated by radiation 
and dark energy. In that case, one would have
\begin{equation}
\rho_{\rm de}\approx {2\over 1-3w_{\rm de}}\rho_{\rm c}(1+z)^2\quad (z\gg z_*)\;,
\end{equation}
and
\begin{equation}
\rho_{\rm r}\approx {3w_{\rm de}+1\over 3w_{\rm de}-1}\rho_{\rm c}(1+z)^2\quad (z\gg z_*)\;,
\end{equation}
implying a relative partitioning of $\rho_{\rm de}\approx 0.8\rho$ and 
$\rho_{\rm r}\approx 0.2\rho$ (if $w_{\rm de}$ continues to be constant
at $-1/2$ towards higher redshifts). But as we say, this discussion is merely 
speculation, and has no bearing on the work reported here. 

\section{Reionization}
We consider reionization from Lyman continuum photons produced in early star-forming galaxies (SFG), 
but note that AGNs may provide part (i.e $\sim 10$ \%) of the reionization (Haardt \& Salvaterra 2015).
The time-dependent cosmic ionization rate in the comoving frame due to star forming galaxies is given by
\begin{equation}
\dot {\bar n} = f_{ion} \,(\eta\,\xi_{ion})\, \rho_{SFR}\;,
\end{equation}
where $f_{ion}$ is the fraction of stellar Lyman continuum photons that escape the galaxy and
reionize the IGM, $\rho_{SFR}$ is the star-formation rate (SFR) density, and $\eta\,\xi_{ion}$
is the number of Lyman continuum photons produced per second per unit SFR 
scaled to the fiducial value 
\begin{equation}
\xi_{ion} = 1.38 \times 10^{53} \,{\rm ph} \,{\rm s}^{-1}\, M_\odot^{-1} \,{\rm yr}\,,
\end{equation}
through the model parameter $\eta$ that takes into account the uncertainty in the photon production efficiency 
(Topping \& Schull 2015). This expression for $\dot {\bar n}$ assumes that all the Lyman continuum photons
escaping into the IGM end up contributing to the reionization. 

We use the empirically derived expression of Robertson et al. (2015) for the star formation rate density 
\be
\rho_{SFR} = a_p \, {(1+z)^{b_p} \over 1+ [(1+z)/c_p]^{d_p}}\;,
\ee
and adopt their best-fit values $a_p = 0.01376$ $M_\odot$ yr$^{-1}$ Mpc$^{-3}$, $b_p = 3.26$, $c_p = 2.59$
and $d_p = 5.68$. It is appropriate for us to do this because, even though the fitting parameters were 
obtained by folding in the optical depth determined with Planck, the fitting values change by less than 
1\% without the Thomson optical depth constraint. Nonetheless, it is important to note that the template 
given in Equation~(23) was optimized for a $\Lambda$CDM-based analysis, and therefore needs to be 
rescaled for the $R_h = ct$ cosmology by the ratio of comoving differential volumes ($1/R_V$) 
in these two models. 

The determination of $\rho_{SFR}$ also relies on a measurement of several luminosity
functions used to estimate the star-formation rate itself. This means that the data 
used to optimize the empirical fit for $\rho_{SFR}$ should be recalibrated for each
individual cosmology. For the specific comparison between $R_{\rm h}=ct$ and the
concordance model, however, the difference in the luminosity distances $d_L$ between 
these two cosmologies is less than $20\%$ (and typically less than $10\%$) over the entire
range of relevant redshifts, extending well past $z=10$. In fact, the two luminosity
distances are equal at $z\sim 8$, as one can see in Fig.~3 of Melia (2015c). Whereas
the angular-diameter and comoving distances, and the time histories, differ considerably 
between these two models, the luminosity distances themselves do not. This $\sim 10-20\%$ 
difference in $d_L$ is well within the uncertainties associated with $\rho_{SFR}$, so 
we will defer the recalibration of the SFR data to future work. But we do include the 
much more important differences that arise between these two models through the 
redshift dependence of their comoving volumes, as discussed above. The original expression
from Robertson et al.~(2015) (solid curve), and the rescaled one for $R_{\rm h}=ct$ (dashed curve), 
are plotted in figure~4. As expected, the larger differential volume in the $R_h = ct$ universe 
at higher redshifts leads to a smaller density of ionizing Lyman continuum photons, which in turn 
acts to increase the time required to reionize the IGM. As we shall see below, however, this effect 
is largely offset by the age differences between the $\Lambda$CDM and $R_h = ct$ cosmologies. 

An additional caveat with the use of Equation~(23) from Robertson et al. (2015)
is that the star-formation rate is still poorly known at very high redshifts. This
expression is an empirical fit to integrated measurements based on several implicit
assumptions concerning high-redshift galaxies. It presumes that the minimum galaxy
luminosity is guessed correctly, and that the stellar populations are constant. 
It also adopts a double power-law ansatz for the history. Substantially different
histories at high redshifts result from the use of similar, though distinct,
techniques (see, e.g., Bouwens et al. 2015; Visbal et al. 2015; Mashian et al.
2015). Such differences suggest that the resulting uncertainty in the
reionization history is at least a factor of two, which can also affect the 
inferred escape fraction to a similar degree.

\begin{figure}
\center{\includegraphics[scale=0.6,angle=0]{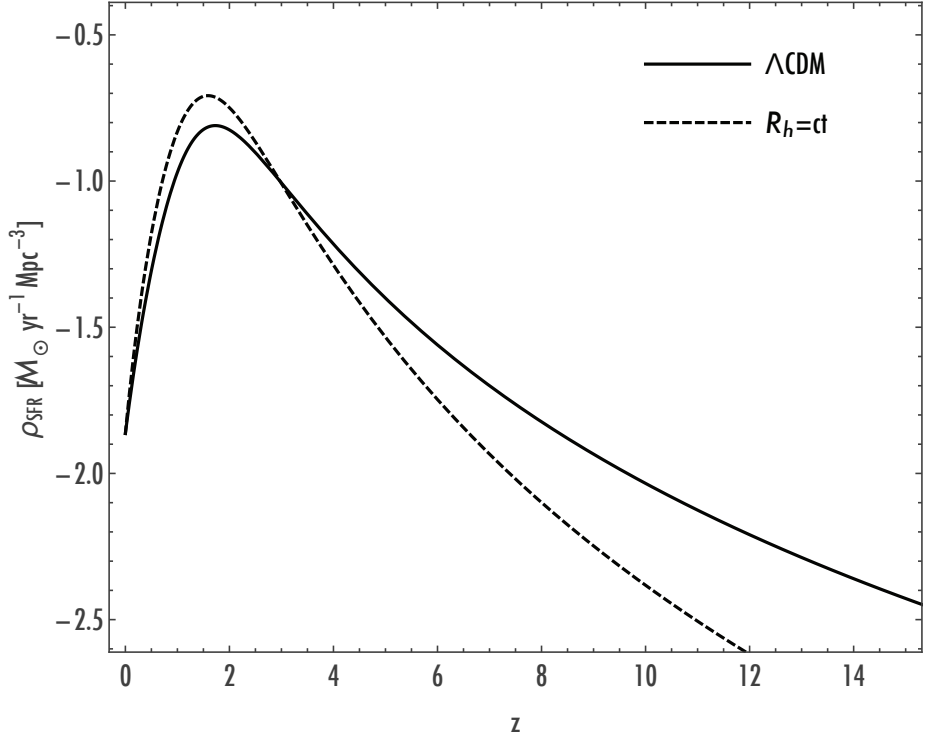}
\caption{The star-formation rate density $\rho_{SFR}$ versus $z$ for $\Lambda$CDM (solid curve) 
and $R_h = ct$ (dashed curve), based on the empirical fit originally published by Robertson et al.~(2015).}} 
\end{figure}

There are two timescales of importance in the reionization process: the characteristic ionization time 
$t_{ion}\equiv \bar n_H / \dot {\bar n}\,,$ and the recombination time $t_{rec} = \left[C_H(z)\, 
\alpha_B(T) \,n_e(z)\right]^{-1}$, written in terms of the clumping factor $C_H$ and recombination
coefficient $\alpha_B$. Note that the comoving hydrogen number density $\bar n_H$ has the same functional 
form, given in Equation~(7), for both $\Lambda$CDM and $R_{\rm h}=ct$. The corresponding proper
electron number density (valid for $z<z_*$, where we assume baryon conservation for both $\Lambda$CDM
and $R_{\rm h}=ct$) is
\be
n_e(z) = f_e\,Q\,{(1-Y_p)\Omega_b\rho_c\over m_H} \, (1+z)^3\,.
\ee 
We emphasize again that the numerical value given in Equation~(7) assumes the fiducial
$\Lambda$CDM constraint $\Omega_b h^2= 0.02230$ for the scaled baryon density, while 
$\Omega_b$ is a free parameter in $R_h = ct$.

Both timescales are fairly well constrained, either through observations or on theoretical grounds, 
though each is subject to some uncertainty. In the case of the ionization time, the production rate of ionizing photons 
depends on metallicity, stellar rotation, and the initial mass function (IMF)---each of which have some 
variability. Various models, based on reasonable assumptions about these characteristics, produce
a Lyman continuum photon production efficiency ranging between $3.22 \times 10^{60}$ and $9.40 \times 
10^{60}$ photons per $M_\odot$ of star formation (Topping \& Shull 2015), corresponding to 
$0.74\le \eta \le 2.2$. As noted in the introduction, the escape fraction of these photons 
has been estimated to be as small as 5\% and as high as 10 - 15\%; values as high as 20\% appear to
be necessary for $\Lambda$CDM in re-ionization studies based on the known population of dwarf galaxies 
at high redshifts (see, e.g., Robertson et al. 2015). Based on current understanding, we therefore 
explore possible outcomes within the range $0.05 \le f_{ion} \le 0.2$. Finally, parameters 
such as $\Omega_b$ have not yet been optimized in the $R_h= ct$ cosmology, though it would be 
reasonable to expect a value ($\lesssim 0.04$) similar to that in the standard model. For this
study, we therefore adopt the range $0.01 \le \Omega_b \le 0.04$. As we shall see, all of
these uncertainties may be incorporated into a single quantity 
\be
{\cal A}\equiv {1\over \eta}\, \left({0.1\over f_{ion}}\right)\,\left({\Omega_b\over 0.02}\right)\,,
\ee
whose expected range is therefore $0.1 \le {\cal A} \le 8.2$. For example, the ionization time 
in the $R_h = ct$ cosmology may then be written 
\be
t^{R_{\rm h}=ct}_{ion} = 0.38\,\,{\cal A}\, R_V(z)\,\left({1+[(1+z)/c_p]^{d_p}\over (1+z)^{b_p}}\right)\,{\rm Gyr}\;.
\ee
The corresponding expression, $t^\Lambda_{ion}$, for $\Lambda$CDM is identical to this, except 
for the omission of the $R_V$ term: $t^\Lambda_{ion}=t^{R_{\rm h}=ct}_{ion}/R_V(z)$.

In addition to the uncertainty in $\Omega_b$ (which appears in the expressions for $\bar{n}_H$ and $n_e$), the 
uncertainties in the recombination time arise from imprecise knowledge concerning the clumpting
factor $C_H$, and the IGM temperature $T$ in the recombination coefficient
\be
\alpha_B = 2.59\times 10^{-13} \, {\rm cm}^3 \, {\rm s}^{-1} \,\left({T\over 10^4 \,{\rm K}}\right)^{-0.845}\,.
\ee
For this study, we adopt the expression from Shull et al. (2012) 
\be
C_H(z) = C_0 \,\left[{(1+z)\over 6}\right]^{-1.1}
\ee
where, following these authors, we take $C_0 = 2.9$ as a fiducial value, but also consider
the range $2\le C_0 \le 10$. Various applications in the literature have considered IGM
temperatures between $5,000$ K and $20,000$ K (see, e.g., Dav\'e et al. 2001; Smith 
et al. 2011), for which $0.56 \le (T/10,000\;{\rm K})^{0.845}\le 1.8$, and we will also
consider this range of values here. These additional uncertainties may be combined into a
second quantity 
\be
{\cal B}\equiv \left({2.9\over C_0}\right) \, \left({T\over 10,000\,{\rm K}}\right)^{0.845}\,
\left({0.02\over \Omega_b}\right)\,,
\ee
which is expected to vary over the range $0.06 \le {\cal B} \le 5.20$. With these definitions,
the recombination time in both $\Lambda$CDM and $R_h = ct$ may be written 
\be
t_{rec} = 70\,\,{{\cal B}\over (1+z)^{1.9}}\,{\rm Gyr}\;,
\ee
assuming $f_e = 1.081$ (for $z>4$, where most of the reionization is thought to have occurred).

\begin{figure}
\center{\includegraphics[scale=0.6,angle=0]{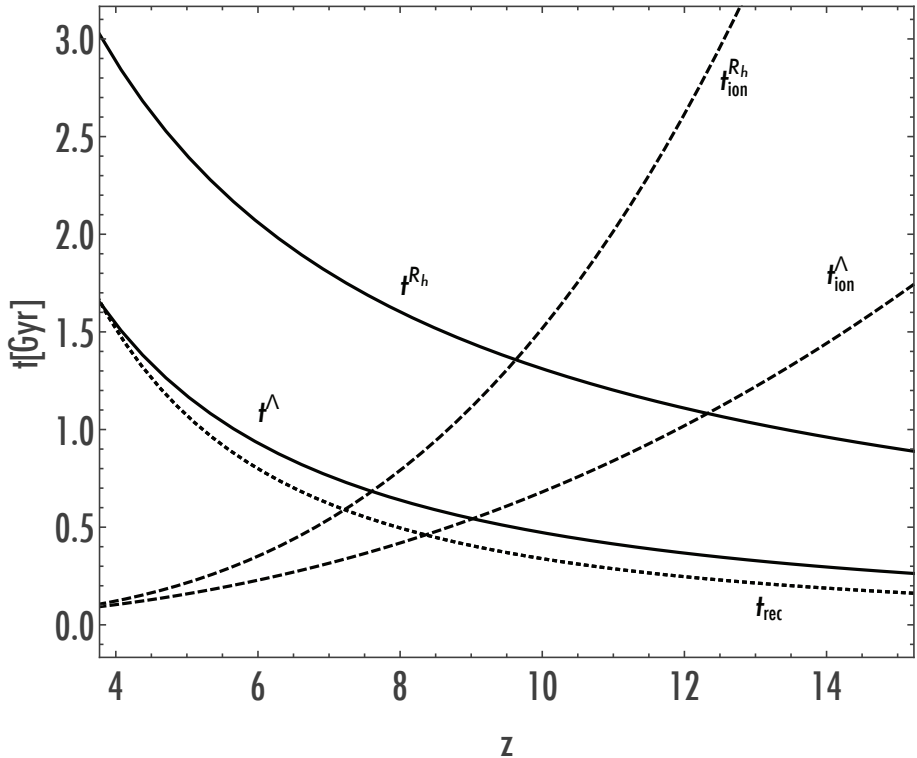}
\caption{The age of the Universe (solid), reionization timescale (dashed), 
and recombination timescale (dotted), as functions of $z$, in $\Lambda$CDM and 
$R_h = ct$, assuming $\eta=1$, $f_{ion} = 0.2$, $T = 20,000$ K, $\Omega_b = 0.0486$, 
and $C_0 = 4.7$. The recombination timescales are identical in these two cosmologies, 
so their curves overlap.}} 
\end{figure}

For illustration, we plot in figure~5 the ionization and recombination times in both 
$R_h = ct$ and $\Lambda$CDM, using parameters similar to those in Robertson et al. (2015),
i.e., $\eta = 1$, $f_{ion} = 0.2$, $T = 20,000$ K and $\Omega_b = 0.02230/h^2 = 0.0486$. 
We mimic the constant clumping factor $C_H = 3$ used by these authors by setting
$C_0 = 4.7$, which then leads to a clumping factor $C_H = 3$ at $z = 8$.\footnote{Note
that the expression for $t_{rec}$ used by Robertson et al. (2015) appears to assume a fully ionized
medium at all redshifts, which enhances the effect of recombination, and therefore 
leads to a slightly later time (i.e., a lower redshift) for reionization to be completed
using our formalism.} This choice of parameters corresponds to ${\cal A} = 1.2$ and ${\cal B} = 0.46$. 

Clearly, for full reionization to occur, the Universe must be older than the reionization 
time ($t > t_{ion}$) which, in turn, must be shorter than the recombination time ($t_{ion} < 
t_{rec}$). In both cosmologies, the latter constraint is realized after the former. For
these illustrative parameter values, the characteristic reionization time therefore
corresponds to $z\approx 8.5$ in $\Lambda$CDM and $z\approx 7.5$ in $R_h = ct$. 
In the next section, we will discuss the detailed solution to the evolution equation for $Q$, 
and compare the results in these two cosmologies.

\begin{figure}
\center{\includegraphics[scale=0.8,angle=0]{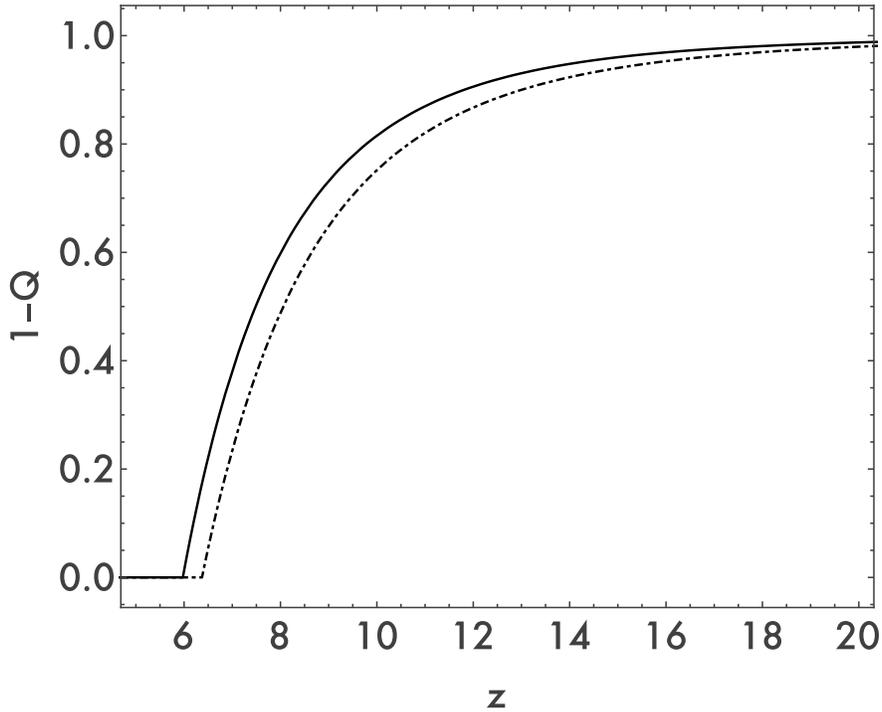}
\caption{Reionization history for ${\cal A} = 1.2$ and ${\cal B} = 0.46$ (see
figure~5): (dashed-dotted) $\Lambda$CDM; (solid) the $R_h = ct$ Universe.}}
\end{figure}

\section{Analysis}
The evolution of the IGM ionization fraction $Q(z)$ is computed by solving the differential
equation
\be
{dQ\over dt} = {\dot {\bar n} \over \bar n_H} - {Q\over t_{rec}}\,,
\ee
assuming specific initial conditions, which include the value $Q = 0$ at $z = 50$ (though we 
note that the results are not sensitive to the initial redshift, as long as it is $> 15$).
To understand how the differences between $\Lambda$CDM and $R_h = ct$ affect the reionization 
history of the Universe, we plot in figure~6 the quantity $1-Q(z)$ versus redshift for
$\Lambda$CDM (dot-dashed) and $R_{\rm h}=ct$ (solid), for the same values ${\cal A} = 1.2$
and ${\cal B} = 0.46$ used to generate figure~5. As expected from our discussion in \S3, 
the reionization rate in $\Lambda$CDM steepens at $z \sim 8.5$. For $R_{\rm h}=ct$, the larger 
differential volume results in a lower star-formation rate density, and hence a longer 
reionization time. This effect delays reionization to lower redshifts. However,
since the $R_{\rm h}=ct$ Universe evolves longer between $z \sim 15$ and $6$
than does the standard model (i.e., 1.16 Gyrs versus 0.66 Gyrs), the volume effect is 
largely offset by the extra time. The net result (for this choice of ${\cal A}$ and
${\cal B}$) is that reionization in $R_{\rm h}=ct$ steepens at $z \sim 7.5$ and the IGM 
becomes fully ionized at a slightly lower redshift than in $\Lambda$CDM. 

\begin{figure}
\center{\includegraphics[scale=0.8,angle=0]{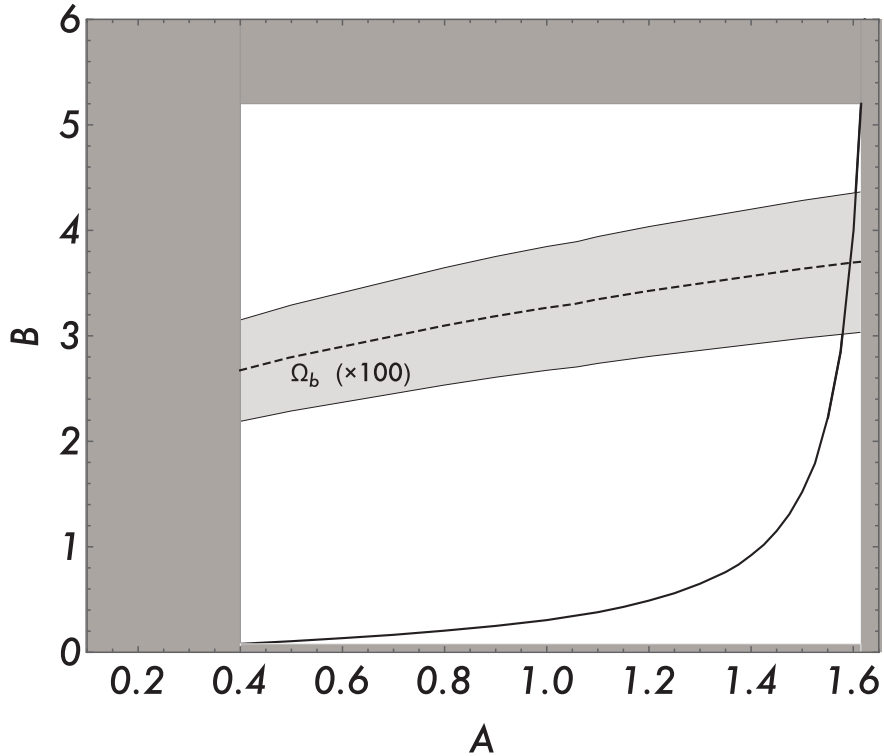}
\caption{(Solid) The value of ${\cal B}$ for a given ${\cal A}$ that allows 
reionization to end by $z = 6$ in the $R_{\rm h}=ct$ Universe. The horizontal
dark shaded regions correspond to values of ${\cal B}$ outside the range 
$(0.06,5.20)$, while the vertical shaded regions exclude values of ${\cal A}$
that similarly require a ${\cal B}$ outside of this range (see text). (Dashed) 
The corresponding value of $\Omega_b$ (as a function of ${\cal A}$) that yields 
an optical depth $\tau = 0.066$ to the scattering surface where the CMB was 
produced (see \S5). The (light shaded) swath surrounding the dashed curve shows 
the uncertainty in $\Omega_b$ corresponding to the possible range in $\tau$, i.e., 
$0.054\lesssim\tau\lesssim 0.078$.}}
\end{figure}

Guided by observations that indicate complete reionization occurs at $z\approx 6$, we next 
optimize the values of ${\cal A}$ and ${\cal B}$ for $R_{\rm h}=ct$ that permit reionization 
to end by this redshift. The results of this ``fitting" are presented in figure~7, where 
the horizontal dark gray shaded areas represent the portion of parameter space outside
the range $0.06 \le {\cal B} \le 5.2$ discussed in \S3. (The vertical dark gray shaded
areas exclude values of ${\cal A}$ that would similarly correspond to ${\cal B}$ outside 
of this range.) And to bracket the possible reionization scenarios in $R_{\rm h}=ct$ for 
all the cases under consideration, we show in figure~8 the reionization histories for 
the extreme values $({\cal A},{\cal B}) = (0.29,0.06)$ and $(1.6,5.2)$. Note that a 
longer recombination time (i.e., a larger value of ${\cal B}$ ) is offset by a longer 
ionization time (i.e., a larger value of ${\cal A}$), and longer timescales delay 
the onset of reionization. In other words, the solid curve in this figure demonstrates
that most of the reionization for the larger values of ${\cal A}$ and ${\cal B}$ occurs
at lower redshifts.

\begin{figure}
\center{\includegraphics[scale=0.8,angle=0]{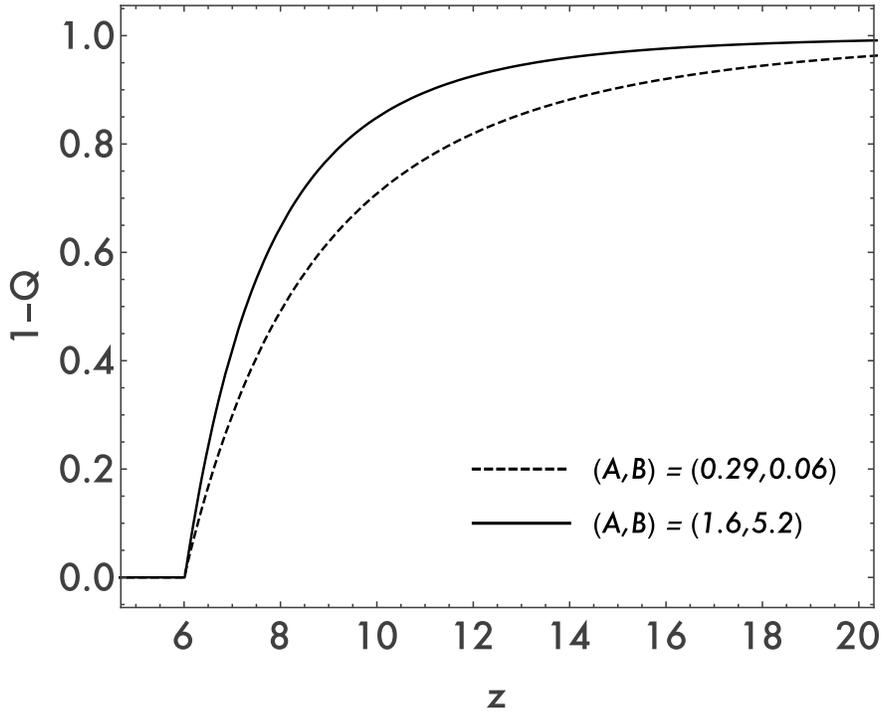}
\caption{Reionization histories in $R_{\rm h}=ct$ for $({\cal A},{\cal B}) = (1.6,5.2)$
(solid) and $(0.52,0.06)$ (dashed). These bracket all the cases considered here that produce
complete reionization by $z = 6$.}}
\end{figure}

\section{Optical Depth Constraints}
In the context of $\Lambda$CDM, the integrated optical depth for Thomson scattering of 
the CMB provides an important constraint on the baryon density and the reionization history. 
The latest results published by the Planck Collaboration (2015) give a value of 
$\tau = 0.066 \pm 0.012$---somewhat lower than the value $\tau = 0.088 \pm 0.14$ 
quoted earlier by the Wilkinson Microwave Anisotropy Probe (Hinshaw et al. 2013). 
According to Robertson et al. (2015), this change in optical depth has resulted
in less tension with other constraints, such as the UV escape fraction $f_{ion}$
and the required number of galaxies at high redshift. 

For the $R_{\rm h}=ct$ cosmology, only the ``low-ell" portion of the CMB spectrum 
has thus far been studied in detail, principally because these moments---corresponding
to angles $>5^\circ-10^\circ$---are influenced most directly by the expansion dynamics
through the Sachs-Wolfe effect (Melia 2014b, 2015a). Conversely, the power spectrum for 
angles $\lesssim 1^\circ$ is generated primarily by local physics, such as the propagation 
of acoustic waves, and is insensitive to the background cosmology (see, e.g., Scott et 
al. 1995). The measurement of the scattering optical depth through the EoR is based
on the interpretation of this power spectrum. Thus, because of this degeneracy
in the ``high-ell" spectrum among different models, and in the absence of a complete analysis 
for the CMB spectrum in $R_{\rm h}=ct$, we will for the time being simply assume that 
the same optical depth constraint may be applied here as in $\Lambda$CDM. 

Starting with the definition $d\tau = \sigma_T\,n_e(z)\,dR$, and using the proper distance
increment $dR = a\,dr = c\,dt = c\,dz (1+z)^{-1}\,H(z)^{-1}$ valid for all cosmologies, one 
easily obtains 
\be
\tau(z) = \sigma_t \,c \int_0^z {n_e(z')\over H(z') (1+z')}\,dz'\,.
\ee
The correct form of $H(z)$ for each individual cosmology must then be used. In the case
of $R_{\rm h} = ct$, the Hubble constant is given by the simple relation $H(z) = H_0 (1+z)$.
And substituting
\be
n_e(z) = f_e(z)\,Q(z)\,n_H(z)
\ee
(where here $n_H(z)$ is the proper hydrogen number density), along with 
\be
n_H(z) = {(1-Y_p)\Omega_b\,\rho_c \over m_H}(1+z)^3\,,
\ee
(with $f_e = 1.081$ when $z > 4$ and $1.162$ for $z\le 4$), we arrive at the expression
\be
\tau = 7.08\times 10^{-4}\,\left({\Omega_b\over 0.02}\right)\,
\int_0^\infty f_e\,Q(z)\,(1+z')\,dz'\,.
\ee
(Note that throughout this paper, we assume $Y_p = 0.2453$ and $h = 0.6774$.)

For a given reionization model, defined by the parameters ${\cal A}$ and
${\cal B}$, one may therefore constrain the fractional baryon density using
the observationally determined optical depth. Doing this for $R_{\rm h}=ct$, 
using the most recent Plank measurement $\tau = 0.066$, we get the $\Omega_b$ 
indicated by the dashed curve in figure~7, as a function of the parameters 
${\cal A}$ and ${\cal B}$ that produce complete reionization by redshift 6. 
The swath bracketing the dashed curve shows the possible uncertainty
in $\Omega_b$ corresponding to the range in optical depth $0.054\lesssim\tau
\lesssim 0.078$. For the parameter values under consideration, we see that 
$\Omega_b$ in this model is restricted to the range $0.026 \lesssim \Omega_b 
\lesssim 0.037$. As expected from the results shown in figure 8, a higher 
baryon density is required to compensate for the later onset of reionization 
that occurs for larger values of ${\cal A}$ and ${\cal B}$. 

\begin{figure}
\center{\includegraphics[scale=0.8,angle=0]{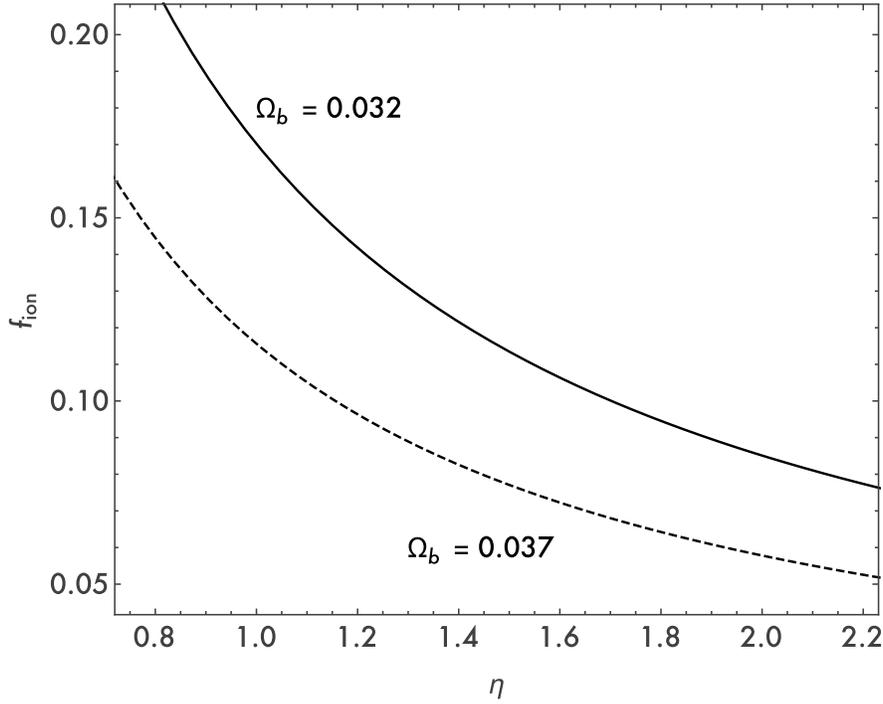}
\caption{The escape fraction $f_{ion}$ as a function of $\eta$, assuming a fractional
baryon density $\Omega_b = 0.032$ (solid curve) and $\Omega_b = 0.037$ (dashed curve)
for parameters that yield complete reionization by redshift
6, and consistent with the optical depth $\tau = 0.066$ measured by Planck.}}
\end{figure}

\section{Discussion}
To see how the observations of complete reionization occurring at $z = 6$ and 
an optical depth of $\tau = 0.066$ constrain the physical parameters under
consideration, we plot in figure~9 the value of the escape fraction $f_{ion}$ as 
a function of $\eta$ and, in figure~10, the value of the clumping factor constant 
$C_0$ as a function of the IGM temperature $T$. The results are bracketed by 
the median value $\Omega_b = 0.032$ (solid curves: ${\cal A} = 0.94$ and 
${\cal B} = 0.27$) and the highest value $\Omega_b = 0.037$ (dashed curves: 
${\cal A} = 1.6$ and ${\cal B} = 5.2$) of the baryon density restricted by
the range of parameters under consideration. Note that a higher baryon 
density, which results from higher values of ${\cal A}$ and ${\cal B}$
(see Figure 7), corresponds to the lower curves in both figures 9 and 10.
The main conclusions drawn from our analysis are as follows:

\begin{figure}
\center{\includegraphics[scale=0.8,angle=0]{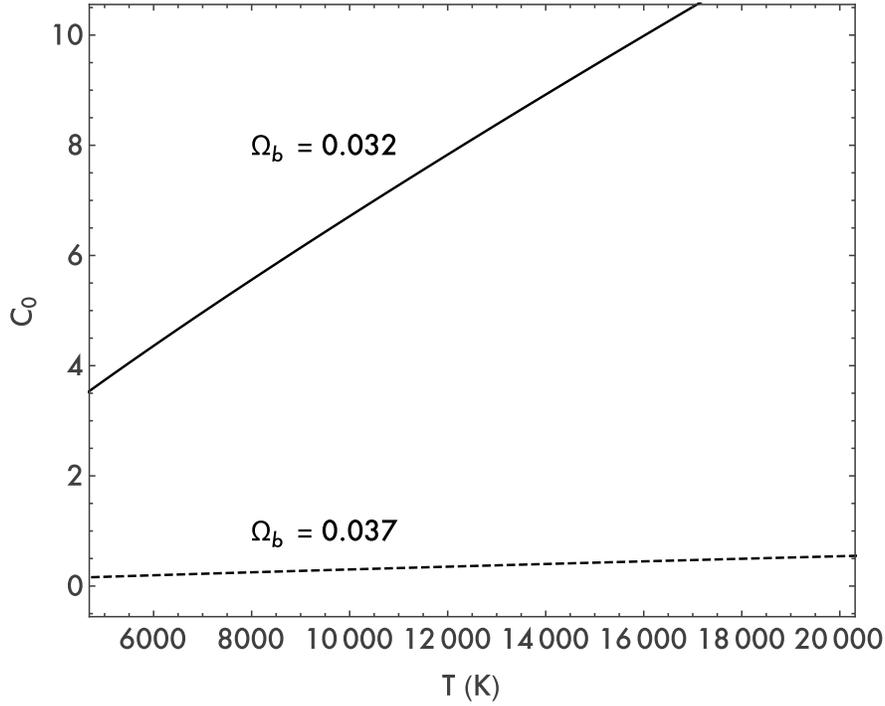}
\caption{The clumping factor constant $C_0$ as a function of IGM temperature 
$T$, assuming a fractional baryon density $\Omega_b = 0.032$ (solid curve) and 
$\Omega_b = 0.037$ (dashed curve) for parameters that yield complete reionization 
by redshift 6, and consistent with the optical depth $\tau = 0.066$ measured by Planck. }}
\end{figure}

1. The $R_{\rm h}=ct$ Universe predicts a different expansion rate and geometry
compared to $\Lambda$CDM, but in spite of these differences, it accounts for
the measured properties of the EoR quite well with physical parameter values
comfortably within their expected ranges.

2. In $R_{\rm h}=ct$, the required escape fraction $f_{ion}$ may be as high
as $\sim 0.2$, as we find in $\Lambda$CDM, but it could also be as small
as $\sim 0.05$. However, to achieve this lower value, one must conclude 
that $\Omega_b=0.037$, $\eta = 2.2$, and that almost no clumping occurs. 

3. Baryon densities lower that $\Omega_b \approx 0.03$ would seem to require large clumping factors. 
It is worth noting that the $R_h = ct$ Universe is older between $z = 6 - 15$ than its $\Lambda$CDM
counterpart, providing more time for clumping to occur.

\section{Conclusion}
As we have alluded to on several occasions, a principal difference between
$\Lambda$CDM and $R_{\rm  h}=ct$ is that the expansion dynamics in the former
is critically dependent on the physical properties of the cosmic fluid, whereas
the expansion dynamics in the latter is strictly fixed by the zero active mass
condition (Melia 2015b; but see also Melia 2007 and Melia \& Shevchuk
2012). Working with $\Lambda$CDM, one is therefore constrained by our
imprecise knowledge concerning, e.g., the equation-of-state of dark energy.
On the flip side, the relative simplicity of the observables in $R_{\rm h}=ct$,
such as the luminosity distance and redshift dependence of the Hubble constant,
has thus far made it unnecessary to study the evolution of its principal 
constituents. For the first time, the present paper addresses this deficiency 
by using measurements of the EoR to examine the evolution of the cosmic fluid 
in this model.

The question of how to account for the measured properties of the EoR has 
been subject to considerable uncertainty, first due to the unknown
sources that actually contribute to the ionizing flux, but more recently due to 
imprecise knowledge concerning how much of this UV radiation finds its way into
the IGM. The latest study by Robertson et al. (2015) suggests that the known
galaxy population out to $z\sim 12$ is sufficient to complete the reionization
process by redshift 6, but only if the UV escape fraction from the higher redshift
dwarf galaxies is $\sim 0.2$, which may be somewhat larger than current estimates
allow.

In this paper, the properties of the EoR have been used to probe the 
evolution of matter, radiation, and dark energy in $R_{\rm h}=ct$---not
just its predicted expansion dynamics. We have argued that, at least out to 
a redshift $\sim 15$, baryons are probably conserved in the comoving frame. 
And with this assumption, we have shown that, consistent with the completion 
of reionization by redshift 6 and a scattering optical depth $\tau\sim 0.066$ 
measured by Planck, the required fractional baryon density falls within the 
``reasonable" range $0.026\lesssim \Omega_b\lesssim 0.037$. 

We have found that, in this model, an escape fraction as low as $\sim 0.05$ 
may be consistent with the measured properties of the EoR, but only if the Lyman
continuum photon production is highly efficient ($\eta = 2.2$) and the baryon
density is at the upper end of its expected range (i.e., $\Omega_b \sim 0.037$).
This additional flexibility compared to $\Lambda$CDM is due to the 
different geometries in these two models (reflected in the differential 
comoving volume), and different time histories (manifested through the $t(z)$
versus $z$ relation).

This interesting result notwithstanding, we should keep in mind the 
important caveat that measurements of $f_{\rm ion}$ and the high-redshift
star-formation rate density are still imprecise. As we have seen, both of
these can significantly affect the reionization history. In galaxies where
$f_{\rm ion}$ can be measured, one reliably finds only upper limits (at
the level of $\sim 5-10\%$). Local galaxies may have even lower escape
fractions than this. It now looks like quasars and AGNs can provide
at most only $\sim 10-20\%$ of the required ionizing photons over the
history of the EoR. The best candidate sources are therefore galaxies
(particularly dwarf galaxies) at high redshift. Unfortunately, estimating
their luminosity function is challenging since it relies heavily on the
assumed minimum luminosity and their evolution with redshift, among
other things. So the analysis presented in this paper should be viewed
as a start of the discussion concerning the EoR in the $R_{\rm h}=ct$
Universe, but much work remains to be done, both observationally and
theoretically. In particular, we point out that the properties of the
EoR presented here are based primarily on the empirically determined 
star-formation rate and galaxy formation and evolution. Large-scale
structure simulations within the $R_{\rm h}=ct$ cosmology have yet
to be completed, so there is no direct evidence that this expansion
scenario can create a population of galaxies consistent with the
observations. The results of this extended investigation are 
forthcoming and will be reported elsewhere.

As of today, the $R_{\rm h}=ct$ cosmology has passed many observational
tests, but almost always based on its global predictions, independent of its 
physical constituents. The fact that this model can also account for the EoR
is an important start to the process of understanding whether or not 
$R_{\rm h}=ct$ is truly a viable representation of cosmic evolution. Future 
work should include an assessment of the fact that baryon conservation
is almost certainly violated at redshifts $z>15$, and the influence of
dark energy does not disappear towards $t=0$, as it does in the standard
model. Baryon number is not conserved in $\Lambda$CDM either, but in this
case, the violation is required only in the first few minutes following
the big bang. In $R_{\rm h}=ct$, on the other hand, the baryon number
in the comoving frame probably changes for a much longer period. In addition,
dark energy cannot be a cosmological constant; it is dynamic, suggesting 
particle physics beyond the standard model. These two features are probably
not independent of each other, but it is still too early for us to know.
\vfill\newpage

\section*{Acknowledgments}
We are very grateful for the thoughtful and helpful comments provided by the 
anonymous referee. F.M. is grateful to Amherst College for its support through 
a John Woodruff Simpson Lectureship, and to Purple Mountain Observatory in Nanjing, 
China, for its hospitality while part of this work was being carried out. This work 
was partially supported by grant 2012T1J0011 from The Chinese Academy of Sciences 
Visiting Professorships for Senior International Scientists, and grant GDJ20120491013 
from the Chinese State Administration of Foreign Experts Affairs. M.F. is supported 
at Xavier University through the Hauck Foundation.

\end{document}